\documentclass{acm_proc_article-sp}

\usepackage{cite}
\usepackage{booktabs}
\usepackage{xspace}
\usepackage{graphicx}
\usepackage{float}
\usepackage{stfloats}

\usepackage{algorithm2e}
\usepackage{algpseudocode}
\usepackage{amsmath}
\usepackage{graphics}
\usepackage{epsfig}
\usepackage{tabularx}
\usepackage{tabu}
\usepackage{booktabs}

\usepackage{multirow}
\usepackage[table,xcdraw]{xcolor}
\usepackage{amsmath}

\usepackage{hyperref}
\usepackage{verbatim}
\usepackage{listings}
\usepackage{graphicx}       % For minipage
\usepackage{flushend}

\newcommand{\tabincell}[2]{\begin{tabular}{@{}#1@{}}#2\end{tabular}} 
\newcommand{\tabfootnote}[1]{\raggedright \scriptsize #1}

\begin{document}

\conferenceinfo{FSE'14}{, November 16--22, 2014, Hong Kong, China}
\CopyrightYear{2014}
\crdata{978-1-4503-3056-5/14/11}

\title{Test Case Purification for Improving Fault Localization}

\numberofauthors{2} 

\author{
\alignauthor Jifeng Xuan\\
       \affaddr{INRIA Lille - Nord Europe}\\
       \affaddr{Lille, France}\\        
       \email{jifeng.xuan@inria.fr}
\alignauthor Martin Monperrus\\
       \affaddr{University of Lille \& INRIA}\\
       \affaddr{Lille, France}\\        
       \email{martin.monperrus@univ-lille1.fr}
}

\date{12 August 2014}

\maketitle
\begin{abstract}
Finding and fixing bugs are time-consuming activities in software development. Spectrum-based fault localization aims to identify the faulty position in source code based on the execution trace of test cases. Failing test cases and their assertions form test oracles for the failing behavior of the system under analysis. In this paper, we propose a novel concept of spectrum driven test case purification for improving fault localization. The goal of test case purification is to separate existing test cases into small fractions (called \textit{purified test cases}) and to enhance the test oracles to further localize faults. Combining with an original fault localization technique (e.g., Tarantula), test case purification results in better ranking the program statements. Our experiments on 1800 faults in six open-source Java programs show that test case purification can effectively improve existing fault localization techniques. 
\end{abstract}

%A category including the fourth, optional field follows...



\section{Introduction}
\label{sect:intro}

Finding and fixing bugs are essential and time-consuming activities in software development. Once a bug is submitted, developers must allocate some effort to identify the exact location of the bug in source code \cite{jones2005empirical}. The problem of localizing bugs in a program is known as \textit{fault localization}, which consists of automatically ranking program entities (e.g., program methods or statements) based on an oracle of the bug, usually a failing test case \cite{zhang2012fault}. \textit{Spectrum-based fault localization} (also known as \textit{coverage-based fault localization}) is a family of methods that use the execution trace of test cases (i.e., the coverage data) to measure the faultyness probabilities of program entities. For example, Tarantula \cite{jones2005empirical}, Jaccard \cite{abreu2007accuracy}, and Ochiai \cite{abreu2007accuracy} are popular spectrum-based fault localization techniques. According to the fault localization rankings, the developers manually examine the program under debugging to find out the location of the bug. 

In modern test-driven software development, unit testing plays an important role for ensuring the quality of software. A unit test framework, such as JUnit for Java, NUnit for .Net, and CPPUnit for C++, provides a platform for developers to manage and automatically execute test cases \cite{meszaros2007xunit}. Each test case is formed as a test method, which employs a test oracle to ensure the expected behavior. The test oracle in a test case is implemented as a set of executable \textit{assertions} for verifying the correctness of the program behavior. For instance, an open source project, Apache Commons Lang (Version 2.6), consists of 1874 test cases with 10869 assertions testing the behavior of over 55K lines of code. That is, each test case includes 5.80 assertions on average. If an assertion in a test case is violated, the unit test framework aborts the execution of this test case and reports the test result (i.e., the test case is failed). 

Test cases can be employed for fault localization \cite{steimann2013threats}, \cite{xu2013general}, \cite{campos2013entropy}. Aborting the execution of a failing test case omits all the unexecuted assertions that are in the same test case. However, the effectiveness of fault localization depends on the quantity of test oracles. \textit{Our key intuition is that recovering the execution of those omitted assertions can lead to more test cases and further enhance the ability of fault localization}. 

In this paper, we propose the concept of spectrum driven test case purification (\textit{test case purification} for short) for improving fault localization. The goal of test case purification is to generate \textit{purified} versions of failing test cases, which include only one assertion per test and excludes unrelated statements of this assertion. We leverage those purified test cases to better localize software faults in Java projects. Test case purification for fault localization consists of three major phases: test case atomization, test case slicing, and rank refinement. First, test case atomization generates a set of single-assertion test cases for each failed test case; second, test case slicing removes the unrelated statements in all the failing single-assertion test cases; third, rank refinement combines the spectra of purified test cases with an existing fault localization technique (e.g., Tarantula) and sorts the statements as the final result.  

We evaluate our work on six real-world open-source Java projects with 1800 seeded bugs. We compare our results with six mature fault localization techniques. Our experimental results show that test case purification can effectively improve the results of existing techniques. Applying test case purification achieves better fault localization on 18 to 43\% of faults (depending on the subject program) and performs worse on only 1.3 to 2.4\% of faults. In terms of fault localization, test case purification on Tarantula (\textit{Tarantula-Purification} for short) obtains the best results among all the techniques we have considered. Tarantula-Purification performs better than Tarantula on 43.28\% of the faults with an average fault-localization improvement of 36.44 statements. With Tarantula-Purification, developers can save half of the effort required for examining faulty statements.  

This paper makes the following major contributions. 

1. We propose the concept of spectrum driven test case purification for improving spectrum-based fault localization. In contrast to novelty in the suspiciousness metric that is common in the fault localization literature, we explore a novel research avenue: the manipulation of test cases to make the best use of existing test data. 

2. We empirically evaluate our approach on 1800 seeded faults on six real-world projects. We compare the fault localization effectiveness of six state-of-the-art techniques (Tarantula, SBI, Ochiai, Jaccard, Ochiai2, and Kulczynski2) with and without test case purification. 

The remainder of this paper is organized as follows. Section \ref{sect:background} presents the background and motivation of our work. Section \ref{sect:purify} proposes the approach to test case purification for improving fault localization. Sections \ref{sect:setup} and \ref{sect:results} show the data sets in the experiments and the experimental results. Section \ref{sect:threats} states the threats to validity in our work. Section \ref{sect:related} lists the related work and Section \ref{sect:conclusions} concludes this paper.

\section{Background and Motivation}
\label{sect:background}

\subsection{Terminology}
\label{subsect:terminology}

We define the major terms used in this paper to avoid ambiguous understanding. 

A \textit{test case} (also called a \textit{test method}) is an executable piece of source code for verifying the behavior of software. In JUnit, a test case is formed as a test method, which consists of two major parts, a test input and a test oracle. A \textit{test input} is the input data to execute the program while a \textit{test oracle} determines the correctness of the software with respect to its test input. Test oracles are created by developers according to business and technical expectations. A test oracle is implemented as a set of executable assertions to ensure that the software performs as expected. A \textit{test suite} is a set of test cases.

An \textit{assertion} is a predicate (a binary expression) that indicates the expected behavior of a program. If an assertion is not satisfied, an exception is thrown. Then the test case containing this assertion aborts and the testing framework reports the failure. For example, \texttt{assertEquals(a, b)} in JUnit is widely used to ensure the equality of values \texttt{a} and \texttt{b}. In practice, a single test case can consist of many assertions (see Section \ref{subsect:subject} for details).

A \textit{subject program} (also called a \textit{proband} \cite{steimann2013threats} or an \textit{object program} \cite{jones2005empirical}) is a program under test. Based on a unit testing framework, like JUnit, a test suite can be automatically executed to test the program. 

A \textit{program entity} represents an analysis granularity for fault localization. For instance, a program entity can be a class, a method, a statement, etc. In this paper, we focus a widely-used program entity, i.e., a statement \cite{jones2005empirical}, \cite{zhang2012fault}, \cite{baah2011mitigating}. 

A \textit{spectrum} of a test case is a set of program entities decorated with execution flags. For a given test case, a \textit{flag} of a program entity indicates whether the test case executes (a.k.a. covers) this particular program entity.

\begin{figure*}[!t]
\centering

\includegraphics[width=1\textwidth]{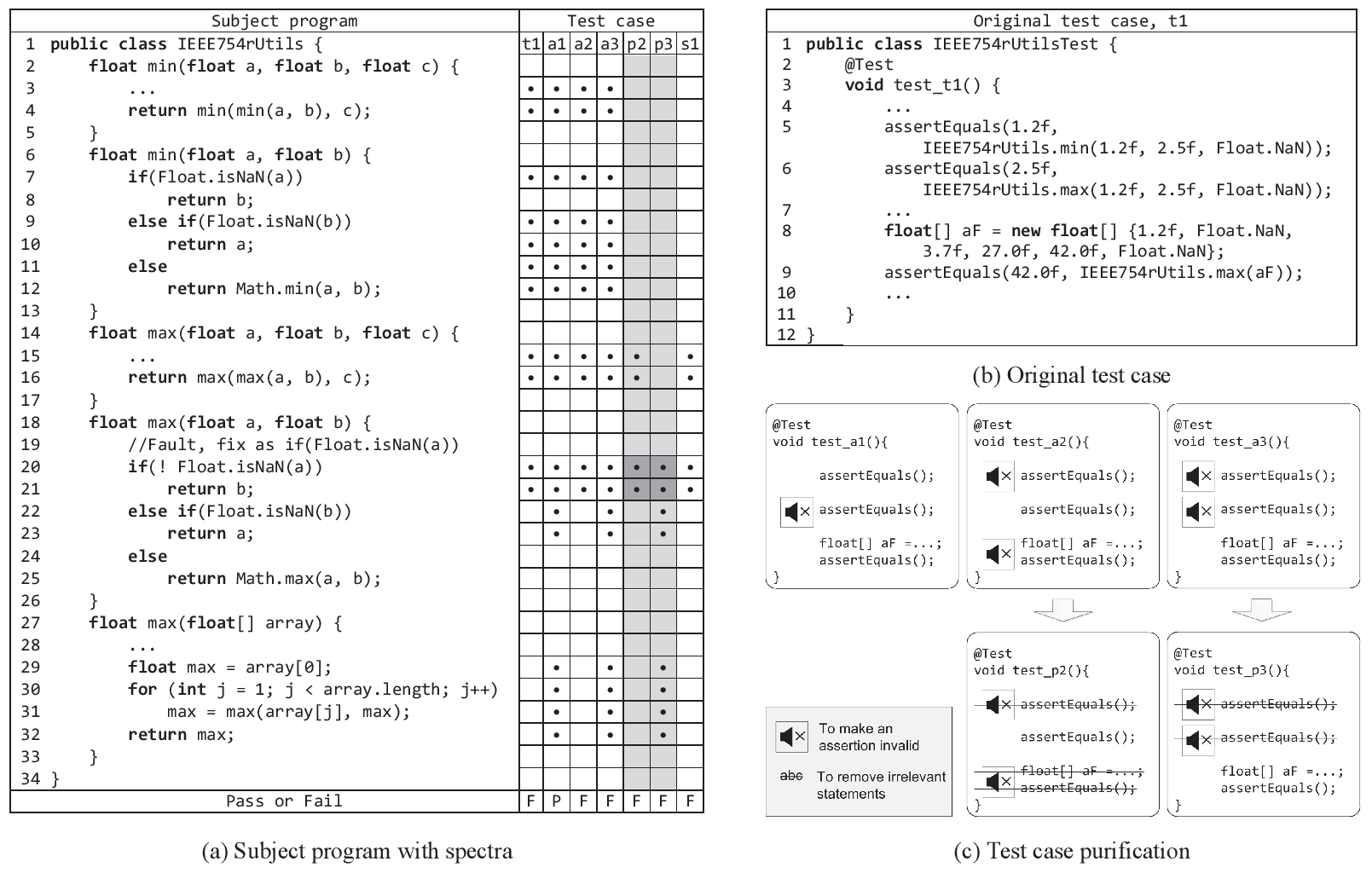}

\caption{Example of test case purification. The subject program and test cases are extracted from Apache Commons Lang 2.6. Test cases \texttt{t1} is the original test cases. Test cases \texttt{a1}, \texttt{a2}, \texttt{a3}, \texttt{p2}, and \texttt{p3} are generated during test case purification. Test case \texttt{s1} is an extra example for the explanation. }
\label{fig:example}
\end{figure*}

In this paper, we focus on subject programs written in Java and tested with JUnit, a unit testing framework. Both JUnit 3 and JUnit 4 are widely used in current Java projects. An intuitive difference between these two versions is that a test case in JUnit 4 starts with a specific annotation \texttt{@Test} and a test case in JUnit 3 is named with a specific convention (in a \texttt{testMethod} style). Our work supports test cases in both  versions of JUnit. Figure \ref{fig:example}(b) briefly illustrates an example of a test case in JUnit 4.

\subsection{Spectrum-Based Fault Localization}
\label{subsect:spectrum}

Spectrum-based fault localization \cite{jones2005empirical}, \cite{abreu2007accuracy}, \cite{xie2013metamorphic} (also known as coverage-based fault localization \cite{steimann2013threats}) is a family of approaches to identifying the exact location of bugs in source code. Popular techniques include Tarantula \cite{jones2005empirical} and Ochiai \cite{abreu2007accuracy}. The input of those approaches is the subject program with its test suite. Spectrum-based fault localization executes the whole test suite and collects the spectrum of each test case. All spectra of test cases form a \textit{spectrum matrix} (also called a \textit{test coverage matrix}) and each element in the matrix indicates whether a test case covers a statement. Based on the spectrum matrix, a fault localization approach calculates the suspiciousness for all statements and ranks them according to their suspiciousness. A detailed description of existing fault localization techniques can be found in Section \ref{subsect:approaches}.

\subsection{Motivation}
\label{subsect:motivation}

Figure \ref{fig:example} shows a fraction of a subject program in Apache Commons (AC) Lang 2.6. AC Lang is an extension library for the Java programming language. Figure \ref{fig:example}(a) lists several methods for the calculation of the maximum and the minimum for IEEE 754 floating-point numbers \cite{zuras2008ieee}. Note that we have omitted the modifiers of methods and several statements to reduce the space. 

We inject a fault at Line 20, i.e., \texttt{if(! Float.isNaN(a))}, by negating the original conditional expression. Then we execute all the test cases of AC Lang and the only failing test case during execution can be found in Figure \ref{fig:example}(b). We call this failing test case \texttt{t1}. The test case \texttt{t1} aborts since the assertion at Line 6 is unsatisfied. In all the test cases of AC Lang, only \texttt{t1} fails and 11 statements from Line 3 to Line 21 are executed by \texttt{t1} (as shown in Figure \ref{fig:example}(a). We use Tarantula \cite{jones2005empirical} as an example of fault localization technique. Based on the actual execution of Tarantula, all the 11 statements executed by \texttt{t1} are ranked with the same suspiciousness. Thus, it is hard to identify the fault at Line 20 from these statements. 

However, \texttt{t1} is aborted at Line 6 and the last assertion at Line 9 has not been executed. Thus, we consider making use of the unexecuted assertion to improve fault localization. As shown in Figure \ref{fig:example}(c), we create three copies (\texttt{a1}, \texttt{a2}, and \texttt{a3}) of \texttt{t1}; for each copy, we force two assertions to not throw an exception even if the assertion is unsatisfied. That is, each of test cases \texttt{a1}, \texttt{a2}, and \texttt{a3}, has only one valid assertion. Then we execute test cases \texttt{a1}, \texttt{a2}, and \texttt{a3}; we find that \texttt{a2} and \texttt{a3} fail at Line 6 and Line 9, respectively (actually, \texttt{a2} in this execution expresses the same behavior as \texttt{t1}). For each of \texttt{a2} and \texttt{a3}, we remove the irrelevant statements to Line 6 and Line 9, respectively; then we get two smaller test cases \texttt{p2} and \texttt{p3}. We execute \texttt{p2} and \texttt{p3} and the spectra are represented as columns in Figure \ref{fig:example}(a) in gray. Based on the spectra of \texttt{p2} and \texttt{p3}, statements at Line 20 and Line 21 are executed twice and six other statements are executed only once. Thus, we can rank the two statements at Line 20 and Line 21 as faulty statement, prior to the other statements. 

The reason for ranking the last two statements is that these statements are the frequently executed ones by failing test cases. In other words, the fault in source code causes the failure of \texttt{p2} and \texttt{p3} and the spectra of \texttt{p2} and \texttt{p3} are different. Thus, the two statements are the most suspicious based on the evidence from the two test cases \texttt{p2} and \texttt{p3}. Moreover, if we directly remove irrelevant statements for the original test case \texttt{t1}, all the dependent statements like Lines 15 and 16 will be kept, as shown in \texttt{s1} in Figure \ref{fig:example}(a). For a large subject program, a large number of dependent statements often interrupt the identification of the fault.     

This example motivates our work, test case purification for fault localization. We use test case purification to generate small fractions of test cases to improve the existing techniques in fault localization.

\section{Test Case Purification}
\label{sect:purify}

In this section, we propose the concept of spectrum driven test case purification (\textit{test case purification} for short) for fault localization. We first present the framework in Section \ref{subsect:framework}. Then we show the details of the three main phases in Sections \ref{subsect:atom}, \ref{subsect:slicing}, and \ref{subsect:refine}, respectively. Finally, we discuss the extensibility of test case purification in Section \ref{subsect:discussion}. 

\subsection{Framework}
\label{subsect:framework}

The main goal of test case purification is to generate purified test cases from each failing test case. A \textit{purified test case} is a short test case with only one assertion and is generated by removing several statements from the original failing test case. We employ such purified test cases to improve existing techniques on fault localization. 
  
\begin{figure*}[!t]
\centering

\includegraphics[width=1\textwidth]{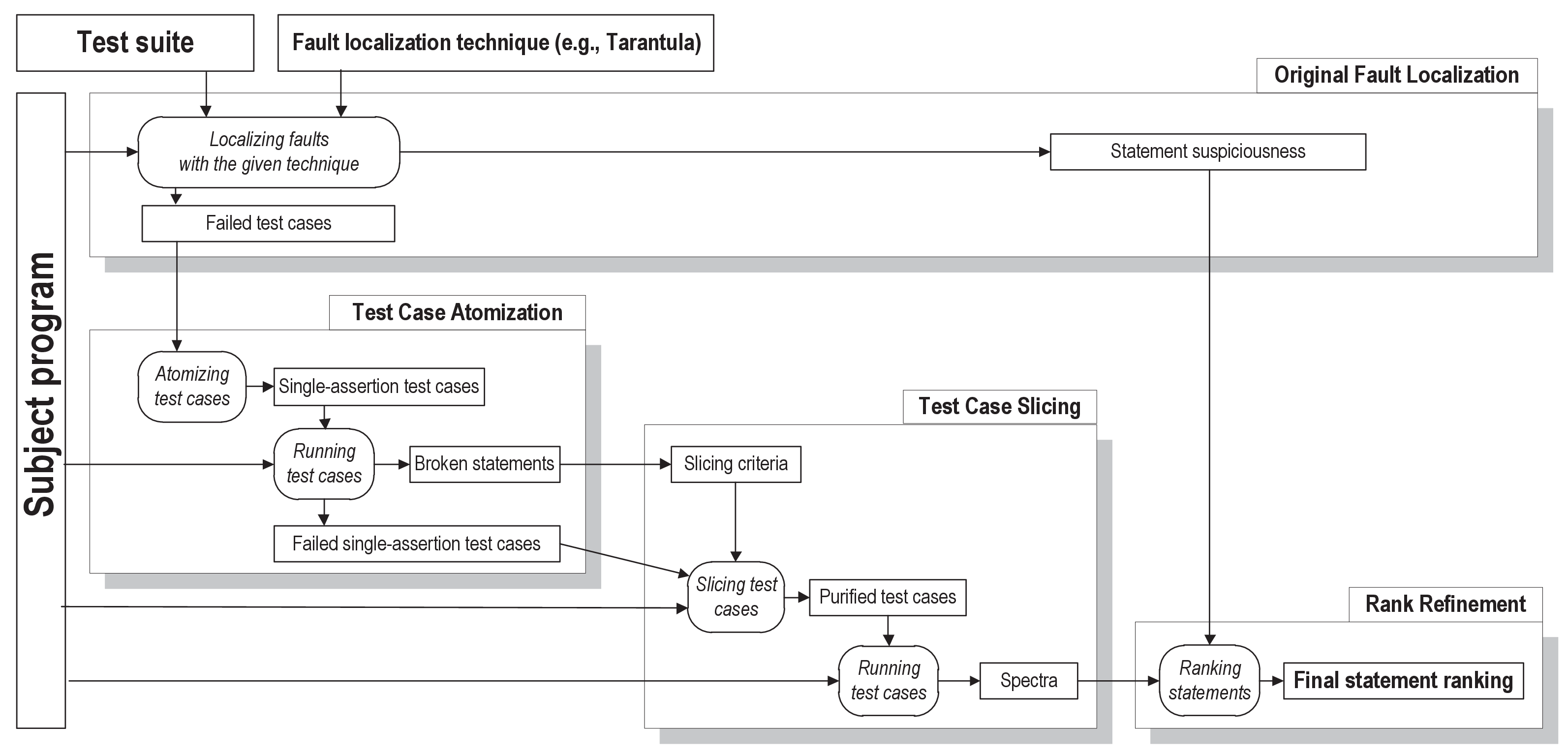}

\caption{Framework of test case purification for fault localization. This framework consists of three phases: test case atomization, test case slicing, and rank refinement. }
\label{fig:framework}
\end{figure*}

Figure \ref{fig:framework} illustrates the framework of test case purification for fault localization. This framework consists of three major phases: test case atomization, test case slicing, and rank refinement. Given a specific technique on fault localization, the input of test case purification is a subject program with its test suite and the final output is a ranking of statements. Both the input and the output are the same as those in typical fault localization techniques, e.g., Tarantula and Ochiai. 

In \textbf{test case atomization}, each original failing test case with $k$ assertions is replaced by $k$ single-assertion test cases. A single-assertion test case is a copy of the original test case, but only one out of $k$ original assertions is kept. In \textbf{test case slicing}, each single-assertion test case is treated as a program. We use dynamic slicing technique to remove irrelevant statements in each single-assertion test case. Then short test cases are generated as purified test cases. In \textbf{rank refinement}, we re-rank the statements in an existing fault localization technique based on the spectra of all the purified test cases. 

In our work, test case purification for fault localization is run automatically. We describe the implementation details in Section \ref{subsect:impl}.

\subsection{Test Case Atomization}
\label{subsect:atom}

The goal of test case atomization is to generate a set of test cases for each failing test case. As the term \textit{atomization} suggests, we consider each assertion has an atomic part in a test case. Given a failing test case with $k$ assertions, we create $k$ copies for this test case and we transform $k-1$ assertions into regular test case statements for each copy (no exception from the assertion reaches the testing framework if the assertion fails). To transform an assertion into a regular test case statement, we surround this assertion with a \texttt{try-catch} structure shown in Figure \ref{fig:surrounding}. 

\begin{figure}[!t]
\centering
%\resizebox{0.5\textwidth}{!}{
%\setlength\tabcolsep{0.6 ex}
%\begin{lstlisting}[frame=single]
\begin{lstlisting}[basicstyle=\ttfamily]  % Start your code-block\\%

  try { 
      /* assertion */ 
  } 
  catch (java.lang.Throwable throwable) {
      /* do nothing */
  }
\end{lstlisting}
%\begin{verbatim}
%try { 
%       /* assertion */ 
%} 
%catch (java.lang.Throwable throwable) {
%       /* do nothing */
%}
%\end{verbatim}
%}
\caption{A surrounding structure for transforming an assertion into a regular test case statement (no exception from the assertion reaches the testing framework if the assertion fails).}
\label{fig:surrounding}
\end{figure}

In Java, the class \texttt{java.lang.Throwable} is a superclass of all the exceptions. As mentioned in Section \ref{subsect:terminology}, an exception will be thrown to the test case if an assertion is not satisfied. Based on the above structure, the exception will be caught as \texttt{throwable} and the test case will not be interrupted\footnote{Sometimes an assertion is originally surrounded by a \texttt{try-catch} statement, e.g., writing files may throw an \texttt{IOException} in Java. Directly adding the surrounding structure to this assertion will cause the compiling error. In this case, we collect the candidate exceptions and add a fraction of dead code to make the compiling pass, like \texttt{if (false) \textbf{throw new} IOException()}.}. Based on the surrounding structure in Figure \ref{fig:surrounding}, a set of $k$ single-assertion test cases are created to replace each originally failing test case. A failing test case with only one assertion will be kept without handling.

Note that in JUnit, two kinds of interruptions will stop the execution of a test case, namely a failure and an error. A \textit{failure} is caused by an unsatisfied assertion, which is designed by developers; an \textit{error} is caused by a fault, which is not considered by developers \cite{gamma2005junit}. Thus, an error may appear in any statement of a test case. In test case atomization, we only deal with the failures (in assertions) in JUnit. If an error appears, the execution of a single-assertion test case will be aborted because an error usually causes severe problems, which are beyond the expected test cases by developers.   

After generating single-assertion test cases, we compile and execute all the single-assertion test cases. Meanwhile, we collect the failing ones among these test cases; for each failing single-assertion test case, we record its position that aborts the execution. This position is referred as a \textit{broken statement}. For example, a broken statement in a single-assertion test case could be an assertion (i.e., the exact assertion left in the test case) or a statement that throws an unexpected error. Finally, each failing single-assertion test case as well as its broken position is collected.   

\subsection{Test Case Slicing}
\label{subsect:slicing}

The goal of test case slicing is to generate purified test cases before collecting their spectra. Given a failing single-assertion test case resulting from test case atomization, we slice this test case by removing irrelevant statements. 

Program slicing can be mainly divided into two categories: static slicing \cite{binkley2007empirical} and dynamic slicing \cite{zhang2007study}. Informally, static slicing keeps all the possible statements based on static data and control dependencies while dynamic slicing keeps the actually executed statements in the dynamic execution (with dynamic data and control dependencies). In test case slicing, we use a dynamic slicing technique to remove statements in test cases since dynamic slicing may lead to more removal of statements \cite{zhang2007study}. In dynamic slicing, a slicing criterion should be specified before execution the program. A slicing criterion is defined as a pair $<b,V>$, where $b$ is a statement in the object program and $V$ is a set of variables to be observed at $b$.

We perform dynamic slicing and slice single-assertion test cases during its execution by the Junit framework. Our slicing criterion for a test case is its broken assertion with all the variables at this statement. Then we execute the dynamic slicing technique to collect the statements that will be removed. After the slicing, each failing single-assertion test case in test case atomization is updated with a purified test case. Then we execute these purified test cases on the project program and record the spectra for next phase. 

\subsection{Rank Refinement}
\label{subsect:refine}

The goal of rank refinement is to re-rank the statements by an existing fault localization technique with the spectra in the phase of test case slicing. 

In all the purified test cases, we keep only one test case if two or more than test cases have the same spectrum. As mentioned in Section \ref{subsect:slicing}, all the purified test cases are failing test cases. Let $S$ be a set of candidate statements. We define the ratio of a statement $s \in S$. First, for a statement $s \in S$ that is covered during the execution of all the purified test cases, $ratio(s)=\frac{\beta_{ef}(s)}{\beta_{ef}(s)+\beta_{nf}(s)}$, where ${\beta_{ef}(s)}$ and ${\beta_{nf}(s)}$ are the numbers of test cases covering and non-covering $s$. Second, for a statement $s$ that is not covered by any purified test case, we directly set $ratio(s)=0$. 

The output of an existing fault localization technique, such Tarantula or Ochiai, is the suspiciousness values for all the candidate statements. Let $susp(s)$ be the suspiciousness value of a statement $s \in S$ in a fault localization technique. Then we normalize the $susp(s)$ as $0$ to $1$ for all the statements in $S$. The normalized suspiciousness value is defined as $norm(s)=\frac{susp(s)-min⁡(S)}{max⁡(S)-min⁡(S)}$, where $min⁡(S)$ and $max⁡(S)$ denote the minimum score and the maximum score for all the statements in $S$, respectively. 

For each statement $s \in S$, both $ratio(s)$ and $norm(s)$ is between $0$ and $1$ (both inclusive). Then we refine the ranking of each statement $s$ by combining $ratio(s)$ and $norm(s)$. The final score of $s$ is defined as $score(s)=norm(s) \times \frac{1+ratio(s)}{2}$. Then for all the statements $s \in S$, the final score $score(s)$ is between $0$ and $1$ (both inclusive). Based on the final scores of all the statements, we re-rank the statements as the result of fault localization by test case purification.  

\subsection{Discussion}
\label{subsect:discussion}

\textbf{Basic fault localization technique}. Test case purification modifies the existing test cases. Consequently, the spectra are changed and the suspicious statements according to a fault localization technique (e.g., Tarantula) are re-ranked. Many other fault localization techniques can be used instead, such as Ochiai, Jaccard, and SBI. We examine the results for six fault localization techniques in Section \ref{sect:results}.

\textbf{Method of rank refinement}. We define the new score of each statement $s$ as $score(s)$. This definition can be replaced by other formulae, for example, the average of $norm(s)$ and $ratio(s)$, i.e., $\frac{norm(s) + ratio(s)} {2}$, or the geometric mean, i.e., $\frac{2 \times norm(s) \times ratio(s)}{norm(s) + ratio(s)}$. Results of such refinement methods can be further explored.

\section{Experimental Setup}
\label{sect:setup}

\subsection{Subject Programs}
\label{subsect:subject}

We select six open-source subject programs for our experiments. Table \ref{tab:subject} gives the key descriptive statistics of those subject programs. All six programs are Java libraries, which are widely used in fault localization research \cite{campos2013entropy}, \cite{inozemtseva2014coverage}, \cite{steimann2012improving}, \cite{steimann2013threats}. We compute the size metric (Source Line of Code - SLoC) with CLOC\footnote{CLOC, \url{http://cloc.sourceforge.net/}.}. The subject programs in our selection are provided with large test suites written in JUnit. For each subject program, we execute the original program with its dependent libraries. We confirm that the whole test suite passes, i.e., our experimental configuration is correct.  

\begin{table*}[!t]
\caption{Subject programs with source code, test suites, and faulty versions}
\label{tab:subject}
\centering
%\begin{minipage}[b]{1\linewidth}
\resizebox{1\textwidth}{!}{
\setlength\tabcolsep{0.4 ex}
\begin{tabular}{|c|cc|cccccc|cc|}
\hline 
\multirow{2}{*}{Subject program} & \multicolumn{2}{c|}{Program source} & \multicolumn{6}{c|}{Test suite}                                                                   & \multicolumn{2}{c|}{Faulty version} \\       \cline{2-11} 
                                 & \#Classes         & SLoC          & \#Classes & SLoC  & \tabincell{c}{JUnit\\ version}   & \tabincell{c}{\#Test\\ cases} & \#Assertions & \tabincell{c}{\#Assertions\\ per test case} & \#Mutants       & \#Faults       \\  \hline\hline
JExel 1.0.0 beta13               & 45                 & 2638          & 43         & 9271  & 4             & 343           & 335           & \ \ 0.98 \dag    & 347              & 313             \\
JParsec 2.0.1                    & 100                & 9869          & 38         & 5678  & 4             & 536           & 869           & 1.62                        & 1698             & 1564            \\
Jaxen 1.1.5                      & 197                & 31993         & 100        & 16330 & 3             & 520           & 585           & 1.13                        & 3930             & 1878            \\
Apache Commons (AC) Codec 1.9    & 56                 & 13948         & 53         & 14472 & 4             & 547           & 1446          & 2.64                        & 2525             & 2251            \\
Apache Commons (AC) Lang 2.6     & 83                 & 55516         & 127        & 43643 & 3             & 1874          & 10869         & 5.80                        & 8830             & 7582            \\
Joda Time 2.3                    & 157                & 68861         & 156        & 69736 & 3             & 4042          & 16548         & 4.09                        & 9197             & 7452            \\             
\hline           
\end{tabular}
}
%\end{minipage}
\tabfootnote{\dag \ In some programs, assertions are abstracted into a specific class, which are not the same assertions in JUnit. In our work, we only handle the assertions in JUnit. Thus, the \# assertions per test case can be less than 1. }
\end{table*}

We follow existing work in \cite{inozemtseva2014coverage}, \cite{steimann2013threats} and use mutation testing tools to create faulty versions. A \textit{mutant} of a program is a copy of the original program with a single change. For instance, a mutant may contain one change of negating a conditional statement. Mutants are meant to simulate likely faults made by developers. Some of mutants (known as \textit{equivalent mutants}) provide the same observable output as the original program. We employ six mutant operators to generate all the mutants for a given subject program. Table \ref{tab:mutant} presents the six mutant operators for generating faulty versions. In our work, we use the PIT tool\footnote{PIT 0.27, \url{http://pitest.org/}.} to generate mutants which has implemented all these six operators. We discard equivalent mutants and keep the faulty versions. Finally, we randomly select 300 mutants from all the seeded faulty versions for each subject program as the final dataset of faulty programs. A thorough study by Steimann et al. \cite{steimann2013threats} has shown that a sample size of 300 mutants gives stable fault localization results.

\begin{table}[!t]
\caption{Mutant operators for generating faulty versions}
\label{tab:mutant}
\centering
\resizebox{0.48\textwidth}{!}{
\setlength\tabcolsep{0.4 ex}
\begin{tabular}{|l|p{20em}|}
\hline
Mutant operator                            & Description                                                    \\  \hline \hline
Invert negatives                           & Invert an integer or a floating-point number as its negative   \\
Return values                              & Change a returned object to null, or increase (or decrease) a returned number  \\
Math                                       & Replace a binary math operator with another math operator      \\
Negate conditionals                        & Negate a condition as its opposite                             \\
Conditional boundary                       & Add or remove the boundary to a conditional statements         \\
Increments                                 & Convert between an increment (++, +=) and a decrement (--, -=)      \\
\hline
\end{tabular}

}
\end{table}

\subsection{Techniques in Comparison}
\label{subsect:approaches}

As explained in Section \ref{sect:purify}, the goal of test case purification is to improve existing fault localization by maximizing the usage of all the assertions. In our experiments, we evaluate the effectiveness of test case purification on six well-studied fault localization techniques: Tarantula, Statistical Bug Isolation (SBI), Ochiai, Jaccard, Ochiai2, and Kulczynski2 (\cite{abreu2007accuracy}, \cite{mao2014slice}, \cite{naish2011model}, \cite{xu2013general}, \cite{zhang2013injecting}). 

Jones et al. \cite{jones2002visualization}, \cite{jones2005empirical} propose Tarantula for fault localization. Tarantula ranks statements by differentiating the execution of failing and passing test cases.  SBI is proposed by Liblit et al. \cite{liblit2005scalable} and calculates the suspiciousness value. Their work shows that the predicted suspicious statements correlate with the root cause. Ochiai is proposed by Abreu et al. \cite{abreu2007accuracy}, which counts both failing test cases and executing test cases. Jaccard is also proposed by Abreu et al. \cite{abreu2007accuracy}. Those four techniques are the most widely-used ones for the evaluation of fault localization. Ochiai2 by Naish et al. \cite{naish2011model} is an extension version of Ochiai; the difference is that Ochiai2 considers the impact of non-executed or passing test cases. Kulczynski2 by Naish et al. \cite{naish2011model} is another widely-used metric. Evaluations of Ochiai2 and Kulczynski2 can be found in \cite{mao2014slice}, \cite{naish2011model}, \cite{xu2013general}.  

Generally, a spectrum-based fault localization technique can be formalized as a formula of calculating the suspiciousness values,
$$susp(s)=f\big(\alpha_{ef} (s),\alpha_{nf} (s),\alpha_{ep} (s),\alpha_{np} (s) \big)$$

\noindent where $\alpha_{ef} (s)$ and $\alpha_{nf} (s)$ are the numbers of failing test cases that execute and do not execute the statement $s$ while $\alpha_{ep} (s)$ and $\alpha_{np} (s)$ are the numbers of passing test cases that execute and do not execute the statement $s$, respectively. Table \ref{tab:technique} summarizes the six techniques that we consider for evaluating test case purification.

\begin{table}[!t]
\caption{Six spectrum-based fault localization techniques in comparison}
\label{tab:technique}
\centering
\resizebox{0.48\textwidth}{!}{
\setlength\tabcolsep{0.4 ex}
\begin{tabular}{|c|c|}
\hline
Technique   & Definition                                                                                                                                                                                                                        \\ \hline\hline
Tarantula   & $\frac {\frac{\alpha_{ef} (s)}{\alpha_{ef} (s)+\alpha_{nf} (s)}} {\frac {\alpha_{ef} (s)} {\alpha_{ef} (s)+\alpha_{nf} (s)} + \frac {\alpha_{ep} (s)} {\alpha_{ep} (s)+\alpha_{np} (s)}} $                                        \\ [2.5ex]
SBI         & $\frac{\alpha_{ef} (s)}{\alpha_{ef} (s)+\alpha_{nf} (s)}$                                                                                                                                                                         \\ [2.5ex]
Ochiai      & $\frac{\alpha_{ef} (s)}{ \sqrt{ \left(\alpha_{ef} (s)+\alpha_{nf} (s) \right) \left(\alpha_{ef} (s)+\alpha_{ep} (s) \right) } }$                                                                                                        \\ [2.5ex]
Jaccard     & $\frac{\alpha_{ef} (s)}{\alpha_{ef} (s)+\alpha_{nf} (s)+\alpha_{ep} (s)}$                                                                                                                                                         \\ [2.5ex]
Ochiai2     & $\frac {\alpha_{ef} (s)  \alpha_{np} (s)} { \sqrt{ \left(\alpha_{ef} (s)+\alpha_{ep} (s)\right)\left(\alpha_{np} (s)+\alpha_{nf} (s)\right)\left(\alpha_{ef} (s)+\alpha_{nf} (s)\right)\left(\alpha_{ep} (s)+\alpha_{np} (s)\right) } } $  \\ [2.5ex]
Kulczynski2 & $\frac{1}{2} \left( \frac{\alpha_{ef} (s)}{\alpha_{ef} (s)+\alpha_{nf} (s)}+\frac {\alpha_{ef} (s)}{\alpha_{ef} (s)+\alpha_{ep} (s)} \right)$    \\ [2.5ex]
\hline                                                                            
\end{tabular}

}
\end{table}

For a given fault localization technique, the \textit{wasted effort} of localizing the faulty statement is defined as the rank of the faulty statement in the ranking according to the suspiciousness values. For statements with the same suspiciousness values, the wasted effort is the average rank between all of them. Formally, the wasted effort of fault localization is defined as 
\begin{tiny}
\vspace{-3ex}
$$\mbox{StmtEffort}=|{s \in S|susp(s)>susp(s^* ) }|+\frac{1}{2} |{s \in S|susp(s)=susp(s^* ) }|+\frac{1}{2}$$
\vspace{-3ex}
\end{tiny}

\noindent where $S$ is a set of candidate statements, $s^* \in S$ is the faulty statement, and $| \cdot |$ indicates the size of a set.

\subsection{Implementation}
\label{subsect:impl}
 
We now discuss the implementation details of our experiment. Our test case purification approach is implemented in Java 1.6. Our experiments run on a machine with an Intel Xeon 2.67 CPU and an Ubuntu 12.04 operating system. Our implementation automatically runs the three phases in Figure \ref{fig:framework}. 

In our work, test suites are automatically executed with Ant 1.8.4\footnote{Ant 1.8.4, \url{http://ant.apache.org/}.} and JUnit 4.11. We set the timeout of running a faulty program as five times of that of the originally correct version to avoid performance bugs \cite{nistor2013toddler}, which may be potentially generated during the program mutation. We execute an existing fault localization technique to compute the original suspiciousness values. We implement the six existing fault localization techniques on top of GZoltar 0.0.3\footnote{GZoltar 0.0.3, \url{http://www.gzoltar.com/}.}. GZoltar \cite{campos2012gzoltar} is a library for facilitating and visualizing fault localization. We use GZoltar to collect the program spectra.

In the phases of test case atomization and test case slicing, we directly manipulate test cases with Spoon 1.5\footnote{Spoon 1.5, \url{http://spoon.gforge.inria.fr/}.}. Spoon \cite{pawlak:inria-00071366} is a library for Java source code transformation and analysis. With the support by Spoon, a Java test class is considered as an abstract syntax tree; and we modify source code via programming abstractions. Spoon also handles annotations in Java hence our implementation fully supports both JUnit 3 and JUnit 4. 

In the phase of test case slicing, we slice test cases with JavaSlicer\footnote{JavaSlicer, \url{https://www.st.cs.uni-saarland.de/javaslicer/}.}. JavaSlicer \cite{hammacher2009profiling} efficiently collects runtime trace for a subject program and removes traces offline with dynamic backward slicing. JavaSlicer requires specifying the point of a thread. Thus, we develop a driver program to facilitate the test case slicing. Since program slicing techniques may cost time and resources, it is necessary to decide how many test cases should be sliced. Based on our experience, it seems that slicing failing test classes one by one is the most efficient, compared to handling failing test cases one by one or all the failing test cases together.

\section{Experimental Results}
\label{sect:results}

In this section, we present our experimental results on test case purification. Section \ref{subsect:overall} presents the overall comparison based on all seeded faults in six subject programs; Section \ref{subsect:detailed} discusses the detailed results for each subject program; Section \ref{subsect:time} evaluates the time cost of test case purification.  

\subsection{Overall Comparison}
\label{subsect:overall}

We compare the capability of our test case purification technique to improve six existing fault localization techniques on six subject programs. 

Table \ref{tab:case} presents the average fault localization results on 1800 seeded faults with mutation. The columns \textit{Positive} gives the absolute and relative numbers of faults, which are improved after applying test case purification, compared to basic techniques in fault localization. Column \textit{Negative} indicates the number of faults when the basic fault localization gives better results. Column \textit{Neutral} shows the number of faults, which are not changed after applying test case purification. 

%\begin{table}[H]
\begin{table}[!t]
%\begin{table}[!hbp]
\caption{Number of faults where test case purification improves existing fault localization techniques (column \textit{Positive}), worsens (column \textit{Negative}) and has no impact (column \textit{Neutral}). Each number is computed over 1800 seeded faults in six subject programs.}
\label{tab:case}
\centering
\resizebox{0.48\textwidth}{!}{
\setlength\tabcolsep{0.4 ex}
\begin{tabular}{|c|cc|cc|cc|}
\hline
\multirow{2}{*}{ \tabincell{c}{Technique\\ in comparison} } & \multicolumn{2}{c|}{Positive} & \multicolumn{2}{c|}{Negative} & \multicolumn{2}{c|}{Neutral} \\  \cline{2-7}
               & \# Faults    & Percent   & \# Faults    & Percent    & \# Faults    & Percent   \\      \hline\hline
Tarantula      & 779          & \textbf{43.28}         & 44           & \textbf{2.44}          & 977          & 54.28        \\
SBI            & 722          & \textbf{40.11}         & 24           & \textbf{1.33}          & 1054         & 58.56        \\
Ochiai         & 373          & \textbf{20.72}        & 28            & \textbf{1.56}         & 1399          & 77.72        \\
Jaccard        & 360          & \textbf{20.00}         & 28           & \textbf{1.56}          & 1412         & 78.44        \\
Ochiai2        & 330          & \textbf{18.33}         & 28           & \textbf{1.56}          & 1442         & 80.11        \\
Kulczynski2    & 666          & \textbf{37.00}         & 24           & \textbf{1.33}          & 1110         & 61.67        \\ 
\hline
\end{tabular}
}
\end{table}

As shown in Table \ref{tab:case}, test case purification improves fault localization for basic fault-localization techniques. For instance, by applying test case purification, 779/1800 (43\%) of faults for Tarantula achieve lower wasted efforts (i.e. faults are easier to be localized and the results are better). The number of faults where purification worsen the ranking is small (worse in 2.44\% of faults for Tarantula), and much smaller than the number of faults that are improved. Except Tarantula, test case purification decreases the effectiveness of fault localization in no more than 28/1800 faults. 

We note that for all the six techniques in our comparison, we obtain neural results on over 50\% of faults. The main reason is that some of the considered faults are easy to localize. For example, for Jaccard, root-cause statements for 389/1800 faults are directly ranked as the first; in those cases, our approach cannot improve the localization since the results are already optimal. Meanwhile, for Jaccard again, 1009/1800 root-cause statements are ranked between the 2nd to the 10th position; and consequently the localization of these faults is hard to improve. In Section \ref{subsect:detailed}, we will show that our test case purification works well for the difficult faults, which are originally localized beyond the top-10 statements.         

Table \ref{tab:effort} presents the wasted effort with or without applying test case purification on 1800 faults. The wasted effort is measured with the absolute number of statements to be examined before finding the faulty one (see Section \ref{subsect:approaches}). It is the main cost of fault localization. In total, there are 12 competing techniques (six fault localization techniques with or without purification). Tarantula with test case purification (called \textit{arantula-Purification} for short) gives the best results among 12 techniques for three of six subject programs. Ochiai-Purification gives the best results for the remaining three subject programs. The last row in Table \ref{tab:effort} gives the average results over all six subject programs. According to this aggregate measure, purification test case improves the wasted effort from 72.06 statements (Tarantula) to 35.62 statements (Tarantula-purification). By applying test case purification with Tarantula, developers save 36.44 statements to examine. In the worst case, they still save 8 statements. 

%\begin{table*}[H]
\begin{table*}[!t]
%\begin{table*}[!hbp]
\caption{Wasted effort (measured with the absolute number of statements to be examined before finding the fault). The wasted effort for our dataset is given on all six considered fault-localization techniques with and without test case purification. The last row averages over all subject programs.}
\label{tab:effort}
\centering
%\begin{minipage}[b]{1\linewidth}
\resizebox{1\textwidth}{!}{
\setlength\tabcolsep{0.6 ex}
\begin{tabular}{|c|cccccc|cccccc|}
\hline
\multirow{2}{*}{ \tabincell{c}{Subject\\ program}} & \multicolumn{6}{c|}{Original technique}                          & \multicolumn{6}{c|}{Test case purification}                    \\ \cline{2-13}
                                 & Tarantula & \ SBI\dag   & Ochiai & Jaccard & \ Ochiai2\dag & Kulczynski2 & Tarantula & SBI    & Ochiai & Jaccard & Ochiai2 & Kulczynski2 \\ \hline\hline
JExel                            & 45.89     & 45.89  & 25.15  & 30.74   & 30.14     & 34.83       & 21.56     & 35.52  & \textbf{21.22}  & 26.90   & 26.98   & 26.96       \\
JParsec                          & 47.76     & 47.76  & 20.67  & 22.46   & 22.46     & 27.02       & \textbf{15.96}     & 18.32  & 20.37  & 21.52   & 21.67   & 17.64       \\
Jaxen                            & 105.88    & 105.88 & 39.02  & 56.38   & 56.38     & 83.46       & 38.92     & 52.01  & \textbf{34.99}  & 45.39   & 46.12   & 70.44       \\
AC Codec                         & 57.04     & 57.04  & 48.27  & 48.53   & 48.53     & 56.25       & \textbf{44.68}     & 49.06  & 46.37  & 47.13   & 47.47   & 48.98       \\
AC Lang                          & 25.66     & 25.66  & 21.92  & 21.99   & 21.99     & 25.61       & 21.24     & 22.12  & \textbf{20.95}  & 20.99   & 21.21   & 22.07       \\
Joda Time                        & 150.13    & 150.13 & 106.35 & 129.03  & 129.03    & 136.53      & \textbf{71.34}     & 104.58 & 101.71 & 124.67  & 124.93  & 100.37      \\	\hline
Average                          & 72.06     & 72.06  & 43.56  & 51.52   & 51.42     & 60.62       & \textbf{35.62}     & 46.93  & 40.93  & 47.77   & 48.06   & 47.74       \\
\hline
\end{tabular}

}
%\end{minipage}
\tabfootnote{\dag \ For some cases, the group of Tarantula and SBI (as well as the group of Ochiai and Ochiai2) produce very similar results. Studies in \cite{liblit2005scalable}, \cite{naish2011model} show evidences on their similarity. The spectra are usually different, which are shown in Table \ref{tab:detail}.}
\end{table*}

\textbf{Summary}. Applying test case purification to the state-of-the-art fault localization techniques results in up to 43\% positive results with the price of 2.4\% worsened faults. Among 12 techniques in comparison, Tarantula-Purification obtains the best results, which are 18.22\% better than the best original technique according to our experimental setup (without purification, Ochiai is the best technique with an average wasted effort of 43.56 statements).

\subsection{Detailed Comparison per Fault Category}
\label{subsect:detailed}

To better understand the effectiveness of test case purification, we analyze all faults in our six subject programs in details. Let $s_{original}$ denotes the original fault localization result, i.e., the wasted effort of localizing the faulty statements as described in Section \ref{subsect:approaches}. We divide the faults in subject programs into three categories according to $s_{original}$, namely faults with $s_{original}=1$, $1<s_{original} \le 10$, and $s_{original}>10$. For example, the faults with $s_{original}=1$ can be viewed as the easy category where there is no space for improving the fault localization since the results are optimal. Similarly, faults with $1<s_{original} \le 10$ can be viewed as the medium category. It is a reasonable task for a developer to examine the top-10 suspicious statements in a program; Le \& Lo \cite{le2013will} also suggest that localizing a fault in top-10 statements is a proof of effectiveness. Results of such faults can be improved a bit. Faults with $s_{original}>10$ can be viewed as representing the hard category. More wasted efforts may need to be checked to localize the faults. 

Table \ref{tab:detail} shows the detailed evaluation on faults in those three categories according to $s_{original}$. Each line is the comparison between an original fault localization technique and test case purification. For each category, we list the positive, negative, neutral (as in Table \ref{tab:case}), and total faults, respectively. We evaluate test case purification with both \# Faults (the number of faults) and StmtSave (the average saved effort obtained by applying test case purification). Note that StmtSave may be below zero because applying test case purification may lead to worse results.

For faults with $s_{original}>10$, applying test case purification can obtain positive and neutral results with few negative results. Taking Tarantula as an example, the effectiveness of fault localization on 524/687 faults (76.27\%) is improved by applying test case purification and worsened for 30 faults (4.37\% in column \textit{Negative}). For Ochiai, localization on 178 out of 394 faults (45.18\%) is improved. Test case purification can save over 65 statements on average for Tarantula or SBI, 36 statements for Kulczynski2, and over 10 statements for Ochiai, Jaccard, or Ochiai2. In JExel, applying test case purification never leads to negative results. That is, fault localization on all the faults with $s_{original}>10$ can be improved or unchanged. In both Jaxen and AC Lang, test case purification can lead to non-negative results on five original techniques except Tarantula (one negative result). In Joda Time, between 11 and 15 cases are worsened. Test case purification in Joda Time performs the worst among our six subject programs. A potential reason is that Joda Time consists of over 68 thousand SLoC; this scale probably hinders fault localization. 

For faults with $1<s_{original} \le 10$, test case purification can also work well. Five out of the six fault localization techniques obtain no more than four negative results; an exception is Tarantula, which obtains 13 negative results. On most of subject programs, applying test case purification can improve the original fault localization, but Tarantula in JParsec as well as Tarantula and Ochiai in Joda Time achieve a little decrease. Note that the result of a fault with $1<s_{original} \le 10$ may not have enough space to improve since the faulty statement has been ranked in top-10 statements. 

For faults with $s_{original}=1$, the results are already optimal. A good approach cannot decrease the results for such faults. In our work, only one fault with $s_{original}=1$ out of 4092 cases in all the subject programs gets a negative result by applying test case purification. In other words, 99.66\% of faults keep an optimal rank under test case purification. 

For all the techniques in our experiments, five out of six subject programs have less than 10 negative results among 300 faults. Joda Time contributes the most negative results, e.g., 20 negatives for Tarantula. On the other side, applying test case purification to Tarantula improves the most among the six original fault localization techniques.   

One major reason for the negative results is that there exists dependency between test cases. For example, if two test cases share a static object and one test case creates the object with a fault (the source code of creating the object contains a faulty statement), then the other test case may fail due to the propagation of the fault. Such propagation makes the second test case fail but the spectrum of the test case does not contain the fault statement. Based on our manual checking, the dependency of test cases is the major reason of negative results. We will further discuss the reasons for negative results in Section \ref{subsect:methodconstr}. 

\textbf{Summary}. Based on the comparison with six techniques on six subject programs, test case purification can improve original techniques in fault localization. For the hard-localized faults (with initial rankings beyond 10 statements), test case purification saves the effort of examining more than 10 statements in average.

\begin{table*}[!t]
\caption{Detailed evaluations on three categories of faults (optimally-localized, easy-localized, hard-localized). Columns \textit{StmtSave} measure the saved effort obtained with test case purification.
}
\label{tab:detail}
\centering
\resizebox{1\textwidth}{!}{
\setlength\tabcolsep{0.3 ex}

\begin{tabular}{|c|c|c||c|c|c|cc||cc|cc|c|cc||cc|cc|c|cc|}
\hline
\multirow{3}{*}{ \tabincell{c}{Subject\\ program} } & \multirow{3}{*}{ \tabincell{c}{Technique in\\ comparison}} & $s_{original}=1$ & \multicolumn{5}{c||}{ $1<s_{original} \le 10$ } & \multicolumn{7}{c||}{ $s_{original}>10$ } & \multicolumn{7}{c|}{ Sum } \\  \cline{3-22}
 &  & Neutral & Positive & Negative & Neutral & \multicolumn{2}{c||}{Total}  & \multicolumn{2}{c|}{Positive}  & \multicolumn{2}{c|}{Negative}  & Neutral & \multicolumn{2}{c||}{Total}  & \multicolumn{2}{c|}{Positive } & \multicolumn{2}{c|}{Negative}  & Neutral & \multicolumn{2}{c|}{Total} \\  \cline{3-22}  
 &  & {\scriptsize \#Faults} & {\scriptsize \#Faults} & {\scriptsize \#Faults} & {\scriptsize \#Faults} & {\scriptsize \#Faults} & {\scriptsize StmtSave} & {\scriptsize \#Faults} & {\scriptsize StmtSave} & {\scriptsize \#Faults} & {\scriptsize StmtSave} & {\scriptsize \#Faults} & {\scriptsize \#Faults} & {\scriptsize StmtSave} & {\scriptsize \#Faults} & {\scriptsize StmtSave} & {\scriptsize \#Faults} & {\scriptsize StmtSave} & {\scriptsize \#Faults} & {\scriptsize \#Faults} & {\scriptsize StmtSave} \\ \hline\hline
\multirow{6}{*}{Jexel} & Tarantula & 50 & 35 & 0 & 104 & 139 & 0.95 & 90 & 79.60 & 0 & 0.00 & 21 & 111 & 64.54 & 125 & 58.37 & 0 & 0.00 & 175 & 300 & 24.32 \\
 & SBI & 50 & 30 & 0 & 109 & 139 & 0.55 & 81 & 37.46 & 0 & 0.00 & 30 & 111 & 27.33 & 111 & 28.02 & 0 & 0.00 & 189 & 300 & 10.37 \\
 & Ochiai & 66 & 13 & 0 & 151 & 164 & 0.11 & 38 & 30.59 & 0 & 0.00 & 32 & 70 & 16.61 & 51 & 23.15 & 0 & 0.00 & 249 & 300 & 3.94 \\
 & Jaccard & 66 & 13 & 0 & 151 & 164 & 0.08 & 38 & 29.96 & 0 & 0.00 & 32 & 70 & 16.26 & 51 & 22.58 & 0 & 0.00 & 249 & 300 & 3.84 \\
 & Ochiai2 & 66 & 10 & 0 & 155 & 165 & 0.06 & 36 & 26.13 & 0 & 0.00 & 33 & 69 & 13.63 & 46 & 20.66 & 0 & 0.00 & 254 & 300 & 3.17 \\
 & Kulczynski2 & 51 & 32 & 0 & 117 & 149 & 0.65 & 73 & 31.01 & 0 & 0.00 & 27 & 100 & 22.64 & 105 & 22.48 & 0 & 0.00 & 195 & 300 & 7.87
\\ \hline\hline
\multirow{6}{*}{Jparsec} & Tarantula & 65 & 36 & 3 & 107 & 146 & -0.09 & 73 & 137.60 & 4 & -123.25 & 12 & 89 & 107.33 & 109 & 92.90 & 7 & -83.93 & 184 & 300 & 31.80 \\
 & SBI & 65 & 29 & 2 & 115 & 146 & 0.22 & 70 & 126.21 & 2 & -18.00 & 17 & 89 & 98.87 & 99 & 89.67 & 4 & -11.63 & 197 & 300 & 29.44 \\
 & Ochiai & 89 & 34 & 0 & 140 & 174 & 0.37 & 11 & 21.55 & 5 & -42.20 & 21 & 37 & 0.70 & 45 & 6.69 & 5 & -42.20 & 250 & 300 & 0.30 \\
 & Jaccard & 89 & 28 & 0 & 140 & 168 & 0.26 & 15 & 20.63 & 5 & -13.90 & 23 & 43 & 5.58 & 43 & 8.20 & 5 & -13.90 & 252 & 300 & 0.94 \\
 & Ochiai2 & 89 & 26 & 0 & 142 & 168 & 0.22 & 14 & 18.96 & 5 & -13.30 & 24 & 43 & 4.63 & 40 & 7.56 & 5 & -13.30 & 255 & 300 & 0.79 \\
 & Kulczynski2 & 73 & 29 & 2 & 120 & 151 & 0.21 & 56 & 50.34 & 2 & -18.00 & 18 & 76 & 36.62 & 85 & 33.66 & 4 & -11.63 & 211 & 300 & 9.38
\\ \hline\hline
\multirow{6}{*}{Jaxen} & Tarantula & 34 & 21 & 0 & 75 & 96 & 0.86 & 148 & 135.29 & 1 & -19.00 & 21 & 170 & 117.67 & 169 & 118.97 & 1 & -19.00 & 130 & 300 & 66.96 \\
 & SBI & 34 & 19 & 0 & 77 & 96 & 0.67 & 149 & 108.03 & 0 & 0.00 & 21 & 170 & 94.68 & 168 & 96.19 & 0 & 0.00 & 132 & 300 & 53.87 \\
 & Ochiai & 63 & 18 & 2 & 138 & 158 & 0.18 & 56 & 21.12 & 0 & 0.00 & 23 & 79 & 14.97 & 74 & 16.53 & 2 & -6.25 & 224 & 300 & 4.04 \\
 & Jaccard & 63 & 13 & 2 & 141 & 156 & 0.19 & 59 & 55.38 & 0 & 0.00 & 22 & 81 & 40.34 & 72 & 45.85 & 2 & -2.25 & 226 & 300 & 10.99 \\
 & Ochiai2 & 63 & 8 & 2 & 146 & 156 & 0.10 & 57 & 53.69 & 0 & 0.00 & 24 & 81 & 37.78 & 65 & 47.38 & 2 & -2.25 & 233 & 300 & 10.25 \\
 & Kulczynski2 & 43 & 31 & 0 & 92 & 123 & 0.86 & 109 & 34.86 & 0 & 0.00 & 25 & 134 & 28.36 & 140 & 27.90 & 0 & 0.00 & 160 & 300 & 13.02
\\ \hline\hline
\multirow{6}{*}{ \tabincell{c}{AC\\ Codec}} & Tarantula & 34 & 36 & 0 & 73 & 109 & 1.23 & 99 & 40.64 & 8 & -56.44 & 50 & 157 & 22.75 & 135 & 30.80 & 8 & -56.44 & 157 & 300 & 12.35 \\
 & SBI & 34 & 26 & 0 & 83 & 109 & 0.83 & 86 & 30.77 & 8 & -43.06 & 63 & 157 & 14.66 & 112 & 24.44 & 8 & -43.06 & 180 & 300 & 7.98 \\
 & Ochiai & 42 & 14 & 0 & 136 & 150 & 0.13 & 29 & 19.53 & 8 & -1.69 & 71 & 108 & 5.12 & 43 & 13.62 & 8 & -1.69 & 249 & 300 & 1.91 \\
 & Jaccard & 42 & 11 & 0 & 139 & 150 & 0.11 & 21 & 20.02 & 8 & -1.81 & 79 & 108 & 3.76 & 32 & 13.64 & 8 & -1.81 & 260 & 300 & 1.41 \\
 & Ochiai2 & 42 & 11 & 0 & 139 & 150 & 0.09 & 18 & 17.81 & 8 & -1.81 & 82 & 108 & 2.83 & 29 & 11.50 & 8 & -1.81 & 263 & 300 & 1.06 \\
 & Kulczynski2 & 34 & 27 & 0 & 86 & 113 & 0.81 & 82 & 29.77 & 8 & -43.94 & 63 & 153 & 13.66 & 109 & 23.23 & 8 & -43.94 & 183 & 300 & 7.27
\\ \hline\hline
\multirow{6}{*}{ \tabincell{c}{AC\\ Lang}} & Tarantula & 43 & 98 & 6 & 104 & 208 & 0.31 & 33 & 38.59 & 2 & -6.50 & 14 & 49 & 25.72 & 131 & 10.83 & 8 & -11.63 & 161 & 300 & 4.42 \\
 & SBI & 43 & 96 & 0 & 112 & 208 & 0.66 & 29 & 31.88 & 0 & 0.00 & 20 & 49 & 18.87 & 125 & 8.50 & 0 & 0.00 & 175 & 300 & 3.54 \\
 & Ochiai & 50 & 92 & 0 & 127 & 219 & 0.55 & 10 & 17.35 & 0 & 0.00 & 21 & 31 & 5.60 & 102 & 2.87 & 0 & 0.00 & 198 & 300 & 0.98 \\
 & Jaccard & 48 & 94 & 0 & 127 & 221 & 0.59 & 10 & 16.95 & 0 & 0.00 & 21 & 31 & 5.47 & 104 & 2.88 & 0 & 0.00 & 196 & 300 & 1.00 \\
 & Ochiai2 & 48 & 89 & 0 & 132 & 221 & 0.52 & 10 & 11.85 & 0 & 0.00 & 21 & 31 & 3.82 & 99 & 2.35 & 0 & 0.00 & 201 & 300 & 0.78 \\
 & Kulczynski2 & 43 & 96 & 0 & 112 & 208 & 0.64 & 29 & 32.09 & 0 & 0.00 & 20 & 49 & 18.99 & 125 & 8.50 & 0 & 0.00 & 175 & 300 & 3.54
\\ \hline\hline
\multirow{6}{*}{ \tabincell{c}{Joda\\ Time} } & Tarantula & \ \ 67 \dag & 29 & 4 & 88 & 121 & -0.14 & 81 & 307.67 & 15 & -92.37 & 15 & 111 & 212.04 & 110 & 227.13 & 20 & -75.30 & 170 & 300 & 78.79 \\
 & SBI & 68 & 27 & 1 & 93 & 121 & 0.33 & 80 & 172.97 & 11 & -27.59 & 20 & 111 & 121.93 & 107 & 129.71 & 12 & -25.42 & 181 & 300 & 45.55 \\
 & Ochiai & 81 & 24 & 2 & 124 & 150 & -0.19 & 34 & 50.96 & 11 & -29.14 & 24 & 69 & 20.46 & 58 & 30.40 & 13 & -29.23 & 229 & 300 & 4.64 \\
 & Jaccard & 81 & 25 & 1 & 124 & 150 & 0.21 & 33 & 52.05 & 12 & -37.63 & 24 & 69 & 18.35 & 58 & 30.17 & 13 & -34.85 & 229 & 300 & 4.35 \\
 & Ochiai2 & 81 & 22 & 1 & 127 & 150 & 0.14 & 29 & 56.95 & 12 & -37.54 & 28 & 69 & 17.41 & 51 & 32.82 & 13 & -34.77 & 236 & 300 & 4.10 \\
 & Kulczynski2 & 70 & 27 & 1 & 95 & 123 & 0.33 & 75 & 149.30 & 11 & -41.86 & 21 & 107 & 100.35 & 102 & 110.19 & 12 & -38.50 & 186 & 300 & 36.16
\\ \hline\hline
\multirow{6}{*}{All} & Tarantula & 293 & 255 & 13 & 551 & 819 & 0.47 & 524 & 128.72 & 30 & -78.73 & 133 & 687 & 94.74 & 779 & 87.41 & 44 & -60.39 & 977 & 1800 & 36.44 \\
 & SBI & 294 & 227 & 3 & 589 & 819 & 0.54 & 495 & 91.66 & 21 & -32.57 & 171 & 687 & 65.05 & 722 & 63.47 & 24 & -29.00 & 1054 & 1800 & 25.12 \\
 & Ochiai & 391 & 195 & 4 & 816 & 1015 & 0.22 & 178 & 28.40 & 24 & -22.71 & 192 & 394 & 11.45 & 373 & 14.33 & 28 & -22.04 & 1399 & 1800 & 2.63 \\
 & Jaccard & 389 & 184 & 3 & 822 & 1009 & 0.26 & 176 & 39.90 & 25 & -21.42 & 201 & 402 & 16.14 & 360 & 20.25 & 28 & -19.34 & 1412 & 1800 & 3.75 \\
 & Ochiai2 & 389 & 166 & 3 & 841 & 1010 & 0.21 & 164 & 38.76 & 25 & -21.26 & 212 & 401 & 14.53 & 330 & 19.92 & 28 & -19.20 & 1442 & 1800 & 3.36 \\
 & Kulczynski2 & 314 & 242 & 3 & 622 & 867 & 0.57 & 424 & 55.31 & 21 & -40.38 & 174 & 619 & 36.52 & 666 & 35.98 & 24 & -35.83 & 1110 & 1800 & 12.87 \\
 \hline
\end{tabular}
}

\tabfootnote{\dag \ Only one fault with $s_{original}$ has a negative result. That is test case purification makes the original result worse on one fault in Joda Time when comparing with Tarantula.}
\end{table*}

\subsection{Computation Time}
\label{subsect:time}

As shown in Table \ref{tab:detail}, Tarantula-Purification obtains the best results among all the techniques in comparison. In this section, we present the computation time of our work. Table \ref{tab:time} lists the computation time of Tarantula-Purification on six subject programs. For each subject program, we list the computation time (in seconds) of the original fault localization and the three phases in test case purification.

The whole process of test case purification costs 275.59 seconds on average. The most time-consuming part is the phase of test case slicing. A major reason for the large computation time is that dynamic program slicing is a complex task and requires monitoring the runtime traces \cite{hammacher2009profiling}, \cite{zhang2007study}. Comparing with the time of original fault localization techniques, i.e., 59.79 seconds, the time of test case purification is still acceptable. We plan to explore further techniques to improve the phase of test case slicing. 

\textbf{Summary}. The computation time of test case purification (275 seconds per fault) is acceptable since the whole process can be executed automatically.

\section{Threats to Validity}
\label{sect:threats}

We discuss threats to the validity of our results with respect to experiment construction and method construction. 

\subsection{Experiment Construction}
\label{subsect:expconstr}

In our work, we evaluate test case purification based on six existing fault localization techniques. Experiments are conducted in six typical open-source subject programs in Java. However, comparing with the large number of existing fault localization techniques, the generality of our work should be further studied.   

To conduct large-scale experiments, we employ mutation testing techniques to inject faults in subject programs. We use six widely-used mutant operators to generate all faulty versions of the subject programs under consideration. Then we randomly select 300 faulty versions as the final faults. A potential threat is that the type of mutant operators may impact the effectiveness of fault localization. For example, a fault localization technique may be good at handling a specific type of faults. We have not checked the results for this issue. We leave it as one of our future work.   

%\begin{table}[H]
\begin{table}[!bhp]
\caption{Time of Tarantula with test case purification (in seconds)}
\label{tab:time}
\centering
\resizebox{0.48\textwidth}{!}{
\setlength\tabcolsep{0.6 ex}
\begin{tabular}{|c||c|ccc||c|}
\hline
\tabincell{c}{Subject\\ program} & \tabincell{c}{Original fault\\ localization} & \tabincell{c}{Test case\\ atomization} & \tabincell{c}{Test case\\ slicing} & \tabincell{c}{Rank\\ refinement} & Total  \\ \hline
JExel           & 5.70                        & 6.54                  & 369.89            & 4.45            & 386.58 \\
JParsec         & 8.11                        & 4.21                  & 137.03            & 5.91            & 155.27 \\
Jaxen           & 14.08                       & 6.38                  & 204.71            & 6.49            & 231.67 \\
AC Codec        & 128.88                      & 41.38                 & 105.60            & 4.20            & 280.06 \\
AC Lang         & 33.31                       & 18.59                 & 79.27             & 9.69            & 140.86 \\
Joda Time       & 168.63                      & 18.06                 & 254.60            & 17.82           & 459.11 \\ \hline
Average         & 59.78                       & 15.86                 & 191.85            & 8.09            & 275.59 \\ 
\hline
\end{tabular}

}
\end{table}

\subsection{Method Construction}
\label{subsect:methodconstr} 

In Section \ref{subsect:refine}, we propose a rank refinement method to leverage the spectra of purified test cases to improve an original fault localization technique. Our method is a simple formula to combine spectra of test case purification and the original fault localization. Other formulae can be used for the combination, e.g., the average and the weighted average. We plan to design new methods to make better use of the spectra of test case purification in the future.

In our work, we generate purified test cases to improve fault localization. Based on the spectra of purified test cases, we rank frequent statements in such spectra prior to other statements. Our experiments show that test case purification can obtain non-negative results on most faults. A potential assumption is that test cases are executed independently. That is, the results of a test case should not impact the results of other test cases. This assumption can be satisfied since most of test cases are well-designed. As mentioned in Section \ref{subsect:detailed}, sometimes test cases suffer from dependencies. This is a challenge topic in fault localization since there is no explicit relationship between the failing test case and the faulty statement.     

We use JavaSlicer \cite{hammacher2009profiling} as the implementation tool in test case slicing. As mentioned by the authors of JavaSlicer, this tool has some known limitations. For example, traces of native methods and Java standard library classes may be missed. To our knowledge, JavaSlicer is the most easy-to-use slicing tool for Java 1.6. In our implementation of test case slicing, we write a program to check potential missing statements by JavaSlicer, but the implementation may still miss some statements.      

\section{Related Work}
\label{sect:related}

To our knowledge, this paper is the first work to directly manipulate test cases to improve fault localization. We list the related work as follows. 

\subsection{Fault Localization Techniques}
\label{subsect:tech} 

Fault localization aims to localize the faulty position in programs. Tarantula by Jones et al. \cite{jones2002visualization} is an integrated framework to localize and visualize faults. Empirical evaluations of Tarantula on fault localization can be found in \cite{jones2005empirical}. Abreu et al. \cite{abreu2007accuracy} propose Ochiai and Jaccard for fault localization. All of Tarantula, Ochiai, and Jaccard can be viewed as the state-of-art in spectrum-based fault localization. Naish et al. \cite{naish2011model} propose a family of fault localization methods and empirically evaluate their results. Recent work by Zhang et al. \cite{zhang2012fault} addresses the problem of how to identify faults with only failed runs. Xie et al. \cite{xie2013theoretical} propose a theoretical analysis on multiple ranking metrics of fault localization and divide these metrics into categories according to their effectiveness.  

Santelices et al. \cite{santelices2009lightweight} combine multiple types of code coverage to find out the faulty positions in program.
%Abreu et al. \cite{abreu2009refining} investigate how to re-rank faulty parts with given program spectra. 
Baah et al. \cite{baah2011mitigating} employ potential outcome model to find out the dynamic program dependencies for fault localization. Xu et al. \cite{xu2013general} develop a noise-reduction framework for localizing Java faults. This work is a general framework that can be used to improve multiple existing fault localization techniques. DiGiuseppe \& Jones \cite{digiuseppe2012semantic} recently propose a semantic fault diagnosis approach, which employs natural language processing to detect the fault locations. Xuan \& Monperrus \cite{xuan2014learning} develop a learning-based approach to combining multiple ranking metrics for fault localizing. Steimann et al. \cite{steimann2013threats} discuss the threats to validity in the empirical assessments of fault localization. Their work also presents the theoretical bounds of the accuracy in fault localization. 
%Recently, Abreu et al. \cite{abreu2009spectrum} and Steimann \& Frenkel \cite{steimann2012improving} have studied the problem of localizing multiple faults, which explores how to apply identify faults in the scenario with multiple faults.  

Hao et al. \cite{hao2010test} propose a test-input reduction approach to reduce the cost of inspecting the test results. Gong et al. \cite{gong2012diversity} design a diversity-maximization-speedup approach to reduce the manual labeling of test cases and improve the accuracy of fault localization. Yoo et al. \cite{yoo2013fault} address the problem of fault localization prioritization. Their work investigates how to rank remaining test cases to maximize fault localization once a previous fault is found.

Baudry et al. \cite{baudry2006improving} leverage the concept of dynamic basic blocks to maximize the ability of diagnosing faults with a test suite.  Artzi et al. \cite{artzi2010directed} directly generate test cases for localizing faults in invalid Html programs in dynamic web applications. This work does not require the test oracles since a web browser can report the crashes once invalid Html programs are found. Fault localization is also used as a phase of predicting a candidate position of the patch in software repair, such as GenProg \cite{le2012genprog} and Nopol \cite{demarco2014automatic}.    

In our work, we address the same problem statement of fault localization. In contrast to existing work, test case purification is a framework to make better use of existing test cases. Our approach directly operates on test cases and can be generally applied to most of existing approaches.

\subsection{Mutation and Slicing Based Fault Localization}
\label{subsect:mutation} 

Mutation-based fault localization has been recently proposed. The kernel idea of mutation-based fault localization is to localize faults by injecting faults. Zhang et al. \cite{zhang2013injecting} propose FIFL, a fault injecting approach to localizing faulty edits in evolving Java programs. Candidate edits are ranked based on the suspiciousness of mutants. Papadakis \& Le Traon \cite{papadakis2013metallaxis} develop Metallaxis-FL, a mutation-based technique for fault localization on C programs. Their work shows that test cases that are able to kill mutants can enable accurate fault localization. Moon et al. \cite{moon2014ask} recently propose MUSE, an approach based on both mutants of faulty statements and mutants of correct statements.  

Slicing-based fault localization leverages program slicing to remove the statements in programs to find out the final faulty statements. Zhang et al. \cite{zhang2007study} employ dynamic slicing to reduce the size of C programs to avoid the distribution by irrelevant statements. Mao et al. \cite{mao2014slice} combine both statistic slicing and dynamic slicing to identify the faulty statements in programs. They empirically evaluate the slicing-based techniques on multiple fault localization techniques. Xie et al. \cite{xie2013metamorphic} propose a new concept of metamorphic slice, based on the integration of metamorphic testing and program slicing. Metamorphic slices localize faults without the requirement of test oracles.

Existing work on mutation-based and slicing-based fault localization aims to change the subject program to identify the faulty parts in the program. In our work, test case purification changes test cases for fault localization rather than subject programs. We make better use of existing test cases (test oracles) to improve the effectiveness of fault localization.

\section{Conclusion}
\label{sect:conclusions}

In this paper, we propose a test case purification approach for improving fault localization. Our work directly manipulates test cases to make better use of existing test oracles. We generate small fractions of test cases, that we call purified test cases, to collect discriminating spectra for all assertions in the test suite under consideration. Our experimental results show that test case purification can effectively improve original fault localization techniques. Only a small fraction of faults (1.3 to 2.4\%) suffer from worsened results. The results show that the benefits of test case purification exist on six fault localization techniques.

As future work, we plan to conduct experiments on other Java projects to further investigate the performance of our work. We plan to design new ranking methods to combine with the spectra of test case purification. Moreover, we want to explore how to reduce the time cost of test case slicing. We plan to check the applicability of the idea of test case purification for other software problems, e.g., regression testing \cite{rothermel2001prioritizing} or automatic software repair \cite{Monperrus2014}.

\section{Acknowledgments}

{\footnotesize
Our work is built on the top of open-source libraries. We thank Renaud Pawlak, Carlos Noguera, and Nicolas Petitprez (for Spoon), Alexandre Perez, Jos\'{e} Carlos Campos, and Rui Abreu (for GZoltar), Clemens Hammacher, Martin Burger, Valentin Dallmeier, and Andreas Zeller (for JavaSlicer) for their contributions.

This research is done with support from EU Project Diversify FP7-ICT-2011-9 \#600654 and an Inria Postdoctoral Research Fellowship.
}

\bibliographystyle{abbrv}

\bibliography{va6d8171f2d864476}  

%\balancecolumns % ACM template, no use.
\flushend

\end{document}